\documentclass[journal]{IEEEtran}
\usepackage{amsmath,amsfonts}
\usepackage{algorithmic}
\usepackage{algorithm}
\usepackage{amssymb}
\usepackage{array}
\usepackage{textcomp}
\usepackage{stfloats}
\usepackage{url}
\usepackage{verbatim}
\usepackage{graphicx}
\usepackage{cite}
\usepackage{makecell}
\usepackage{subcaption}
\usepackage{multirow}
\usepackage[utf8]{inputenc}
\usepackage[T1]{fontenc}
\usepackage{amsthm}
\PassOptionsToPackage{table,xcdraw}{xcolor}
\usepackage{xcolor}
\usepackage{acronym}
\usepackage{xspace}
\usepackage[colorlinks,urlcolor=blue,linkcolor=blue,citecolor=blue]{hyperref}
\usepackage{todonotes}

\def\contd{\textit{containerd}\xspace}
\def\crictl{\texttt{crictl}\xspace}

\def\maximgp{\texttt{MaxParallelImagePulls}\xspace}
\def\mp{$mp$\xspace}
\def\force{\texttt{---force}\xspace}
\def\gperiod{\texttt{---grace-period=0}\xspace}
\def\pod{Pod\xspace}
\def\pods{Pods\xspace}
\def\criapi{\acs{cri}-\acs{api}\xspace}

\begin{document}

\def\papertitle{Exploiting Kubernetes' Image Pull Implementation to Deny Node Availability} 

\title{\papertitle}
\author{
    \IEEEauthorblockN{Luis Augusto Dias Knob\IEEEauthorrefmark{1}, Matteo Franzil\IEEEauthorrefmark{2}\IEEEauthorrefmark{1}, Domenico Siracusa\IEEEauthorrefmark{2}\IEEEauthorrefmark{1}}\\\vspace{10pt}
    \IEEEauthorblockA{\IEEEauthorrefmark{1}Center for Cybersecurity, Fondazione Bruno Kessler
    \\l.diasknob@fbk.eu}\\
    \IEEEauthorblockA{\IEEEauthorrefmark{2}Department of Information Engineering and Computer Science, University of Trento
    \\\{matteo.franzil, domenico.siracusa\}@unitn.it}
}

\newcommand{\reviewerAt}[1]{}
\newcommand{\reviewerAc}[1]{{\color{black}{#1}}}
\newcommand{\reviewerBt}[1]{}
\newcommand{\reviewerBc}[1]{{\color{black}{#1}}}
\newcommand{\reviewerCt}[1]{}
\newcommand{\reviewerCc}[1]{{\color{black}{#1}}}
\newcommand{\reviewerDt}[1]{}
\newcommand{\reviewerDc}[1]{{\color{black}{#1}}}

\newcommand{\generic}[1]{{\color{black}{#1}}}


\markboth{}{}

\maketitle

\acrodef{abac}[ABAC]{Attribute-Based Access Control}
\acrodef{acl}[ACL]{Access Control List}
\acrodef{ahd}[AHD]{Abnormal Health Detection}
\acrodef{ai}[AI]{Artificial Intelligence}
\acrodef{alu}[ALU]{Arithmetic Logic Unit}
\acrodef{ann}[ANN]{Artificial Neural Network}
\acrodef{api}[API]{Application Programming Interface}
\acrodef{aks}[AKS]{Azure Kubernetes Service}
\acrodef{gke}[GKE]{Google Kubernetes Engine}
\acrodef{eks}[EKS]{Amazon Elastic Kubernetes Service}
\acrodef{bmv2}[BMv2]{Behavioral Model version 2}
\acrodef{bnn}[BNN]{Binary Neural Network}
\acrodef{bow}[BoW]{Bag-of-Words}
\acrodef{caas}[CaaS]{Container as a Service}
\acrodef{cicd}[CI/CD]{Continuous Integration/Continuous Delivery}
\acrodef{cncf}[CNCF]{Cloud Native Computing Foundation}
\acrodef{cni}[CNI]{Container Network Interface}
\acrodef{cnn}[CNN]{Convolutional Neural Network}
\acrodef{cpe}[CPE]{Customer Premise Equipment}
\acrodef{cri}[CRI]{Container Runtime Interface}
\acrodef{csi}[CSI]{Container Storage Interface}
\acrodef{ddos}[DDoS]{Distributed Denial of Service}
\acrodef{dlp}[DLP]{Data Loss/Leakage Prevention}
\acrodef{dl}[DL]{Deep Learning}
\acrodef{dnn}[DNN]{Deep Neural Network}
\acrodef{dns}[DNS]{Domain Name System}
\acrodef{dos}[DoS]{Denial of Service}
\acrodef{dpi}[DPI]{Deep Packet Inspection}
\acrodef{ebpf}[eBPF]{extended Berkeley Packet Filter}
\acrodef{ewma}[EWMA]{Exponential Weighted Moving Average}
\acrodef{faas}[FaaS]{Function as a Service}
\acrodef{fl}[FL]{Federated Learning}
\acrodef{fnr}[FNR]{False Negative Rate}
\acrodef{foss}[FOSS]{Free and Open-Source Software}
\acrodef{fpr}[FPR]{False Positive Rate}
\acrodef{fpu}[FPU]{Floating Point Unit}
\acrodef{fvg}[\textsc{FedAvg}]{Federated Averaging}
\acrodef{gc}[GC]{Garbage Collector}
\acrodef{gdpr}[GDPR]{General Data Protection Regulation}
\acrodef{gpu}[GPU]{Graphics Processing Unit}
\acrodef{grpc}[gRPC]{gRPC Remote Procedure Call}
\acrodef{ha}[HA]{Hardware Appliance}
\acrodef{iaas}[IaaS]{Infrastructure as a Service}
\acrodef{ics}[ICS]{Industrial Control System}
\acrodef{ids}[IDS]{Intrusion Detection System}
\acrodef{iid}[i.i.d.]{independent and identically distributed}
\acrodef{ilp}[ILP]{Integer Linear Programming}
\acrodef{iot}[IoT]{Internet of Things}
\acrodef{ips}[IPS]{Intrusion Prevention System}
\acrodef{isp}[ISP]{Internet Service Provider}
\acrodef{jsd}[JSD]{Jensen-Shannon Distance}
\acrodef{k8s}[K8s]{Kubernetes}
\acrodef{kep}[KEP]{Kubernetes Enhancement Proposals}
\acrodef{ldap}[LDAP]{Lightweight Directory Access Protocol}
\acrodef{lstm}[LSTM]{Long Short-Term Memory}
\acrodef{lucid}[\textsc{Lucid}]{Lightweight, Usable CNN in DDoS Detection}
\acrodef{mau}[MAU]{Match-Action Unit}
\acrodef{mbgd}[MBGD]{Mini-Batch Gradient Descent}
\acrodef{mips}[MIPS]{Millions of Instructions Per Second}
\acrodef{mlp}[MLP]{Multi-Layer Perceptron}
\acrodef{ml}[ML]{Machine Learning}
\acrodef{mssql}[MSSQL]{Microsoft SQL}
\acrodef{nat}[NAT]{Network Address Translation}
\acrodef{netbios}[NetBIOS]{Network Basic Input/Output System}
\acrodef{nfv}[NFV]{Network Function Virtualization}
\acrodef{nf}[NF]{Network Function}
\acrodef{nic}[NIC]{Network Interface Controller}
\acrodef{nids}[NIDS]{Network Intrusion Detection System}
\acrodef{nn}[NN]{Neural Network}
\acrodef{nsc}[NSC]{Network Service Chaining}
\acrodef{ntp}[NTP]{Network Time Protocol}
\acrodef{of}[OF]{OpenFlow}
\acrodef{ood}[o.o.d.]{out-of-distribution}
\acrodef{oom}[OOM]{Out-of-Memory}
\acrodef{os}[OS]{Operating System}
\acrodef{ourtool}[P4DDLe]{P4-empowered Ddos detection with Deep Learning}
\acrodef{p4}[P4]{Programming Protocol-independent Packet Processors}
\acrodef{pess}[PESS]{Progressive Embedding of Security Services}
\acrodef{pisa}[PISA]{Protocol Independent Switch Architecture}
\acrodef{pop}[PoP]{Point of Presence}
\acrodef{portmap}[Portmap]{Port Mapper}
\acrodef{ppv}[PPV]{Positive Predictive Value}
\acrodef{pss}[PSS]{Pod Security Standard}
\acrodefplural{pss}[PSSs]{Pod Security Standards}
\acrodef{ps}[PS]{Port Scanner}
\acrodef{qoe}[QoE]{Quality of Experience}
\acrodef{qos}[QoS]{Quality of Service}
\acrodef{rbac}[RBAC]{Role-Based Access Control}
\acrodef{rnn}[RNN]{Recurrent Neural Network}
\acrodef{rpc}[RPC]{Remote Procedure Call}
\acrodef{sa}[SA]{Service Account}
\acrodef{sdn}[SDN]{Software Defined Networking}
\acrodef{sfnr}[sFNR]{system False Negative Rate}
\acrodef{sla}[SLA]{Service Level Agreement}
\acrodef{snf}[SNF]{Security Network Function}
\acrodef{snmp}[SNMP]{Simple Network Management Protocol}
\acrodef{ssdp}[SSDP]{Simple Service Discovery Protocol}
\acrodef{svm}[SVM]{Support Vector Machine}
\acrodef{tc}[TC]{Traffic Classifier}
\acrodef{ttl}[TTL]{Time to Live}
\acrodef{tftp}[TFTP]{Trivial File Transfer Protocol}
\acrodef{tor}[ToR]{Top of Rack}
\acrodef{tpr}[TPR]{True Positive Rate}
\acrodef{tsp}[TSP]{Telecommunication Service Provider}
\acrodef{unb}[UNB]{University of New Brunswick}
\acrodef{vm}[VM]{Virtual Machine}
\acrodef{vnep}[VNEP]{Virtual Network Embedding Problem}
\acrodef{vne}[VNE]{Virtual Network Embedding}
\acrodef{vnf}[VNF]{Virtual Network Function}
\acrodef{vpn}[VPN]{Virtual Private Network}
\acrodef{vsnf}[VSNF]{Virtual Security Network Function}
\acrodef{waf}[WAF]{Web Application Firewall}
\acrodef{wan}[WAN]{Wide Area Network}
\acrodef{xdp}[XDP]{eXpress Data Path}


\begin{abstract}
\ac{k8s} has grown in popularity over the past few years to become the \textit{de-facto} standard for container orchestration in cloud-native environments. While research is not new to topics such as containerization and access control security, the \ac{api} interactions between \ac{k8s} and its runtime interfaces have not been studied thoroughly.
In particular, the \acs{cri}-\acs{api} is responsible for abstracting the container runtime, managing the creation and lifecycle of containers along with the downloads of the respective images. However, this decoupling of concerns and the abstraction of the container runtime renders \ac{k8s} unaware of the status of the downloading process of the container images, obstructing the monitoring of the resources allocated to such process. In this paper, we discuss how this lack of status information can be exploited as a Denial of Service attack in a \ac{k8s} cluster. We show \generic{how} such attacks can \generic{impact worker nodes,} generating up to 95\% average CPU usage, prevent \generic{downloads of} new container images, and increase I/O and network usage for a potentially unlimited amount of time.
\reviewerBc{We argue that solving this problem would require a radical architectural change in the relationship between \ac{k8s} and the \acs{cri}-\acs{api}, which would be unfeasible in the short term. Thus, as a stopgap solution, we propose MAGI: an eBPF-based, proof-of-concept mitigation that detects and terminates potential attacks.}

\end{abstract}


\section{Introduction}
\label{sec:introduction}


Since its inception, \acf{k8s} has been conceived as a container management system to orchestrate multiple workloads in a scalable, resilient, and reliable way \cite{burns_borg_2016}. This containerization deployment-oriented approach enables cluster and application automation, multi-tenancy, and reproducibility. Additionally, \acs{k8s}' loosely coupled architecture based on asynchronous communication, isolation, and decoupling of concerns pioneered a remarkable change in the cloud-native computing landscape.


Being implemented with a modular structure, \ac{k8s} uses a set of \acp{api} to manage the network, storage, and container runtimes. This creates a scenario where several modules with distinct implementation and requirements need to communicate to execute a task (e.g., the creation of containers), usually in an asynchronous way. Such a design provides high availability, scalability, and performance but creates gaps that could potentially be used as attack vectors.

\looseness=-1
In this paper, we investigate the hazards of the asynchronous communications between two modules inside the \ac{k8s} cluster, the Kubelet, the module that runs on each node and communicates with the central API, and the container runtime, which is tasked with handling the creation, management, and deletion of containers in \ac{k8s} nodes. Our study figures that since the communication between these modules happened asynchronously, it does not maintain any intermediate status on each request, enabling a malicious user to exploit this behavior as a threat that can be used as a vector to several distinct attacks.

In particular, when deploying a \pod \xspace --- a group of one or more containers --- an image pull is usually triggered for the corresponding containers if unavailable locally. If a container is deleted before the image is fully downloaded, the \ac{api} will not communicate this action to the container runtime. However, the runtime will still download the image, wasting resources in the process. Similarly, when a container is forcibly destroyed, its resources are not instantly released. While \ac{k8s} will delete all information from the \ac{api} as soon it receives the request, the complete elimination of the containers composing the \pod will only occur when all modules finish their deletion processes. Although this is intended behavior, such a process could be exploited to mount a \ac{dos} attack on the \reviewerBc{cluster's worker nodes}. Indeed, these image download requests would prevent other users from scheduling new jobs without a cached image and force the tenant to consume unnecessary resources.

To understand the actual impact, we present several possible attack scenarios, studying the impact on the scheduling, system resources, and other containers running in the node. We also describe in detail the threat, with its limitations and requirements. We show how this issue affects all Kubernetes deployments, demonstrating its effect both in private and public cloud settings. Additionally, we discuss how this could be solved in \ac{k8s}. Since the solution is not trivial and will require time and effort to be achieved, we present a temporary, eBPF-based proof-of-concept solution to address the problem until the fix is fully implemented in all the relevant projects. 


We also argue that the majority of \ac{k8s} security research over the last ten years focused on lateral container movement, container escapes, and authentication and access control\cite{rahman_security_2023,he_cross_2023, xiao_attacks_2023, almaraz-rivera_anomaly-based_2023,carrion_kubernetes_2023, minna_understanding_2021}. These topics have been extensively researched, and several ideas have been proposed to address these issues. However, studying the interactions between \ac{k8s} and its runtime interfaces is one area that is frequently disregarded both in the literature and in practice, mainly due to the challenges presented by a modular project with several distinct working groups. Taking that into account, the contributions of this work are summarized as follows.

\textbf{New resource-based \ac{dos} threat in \ac{k8s}.} We demonstrate that it is possible to exploit a series of design flaws in the \acl{k8s} \criapi to execute a \ac{dos} attack. We also discuss the importance of \ac{api} status routes and how the asynchronous communication between distinct modules can be used as a vulnerability point.

\looseness=-1
\textbf{Attacking strategies}. We explore how such an attack can overcome a cluster's defenses and mount a \ac{dos} attack on a target node, blocking the image download of other applications and hampering the resource use by the containers. We also show that it is possible to manipulate the node's image cache, increasing the time needed to start applications in the cluster. 

\textbf{\acs{ebpf} proof-of-concept solution}. We discuss how the problem can be mitigated, with the pros and cons of each solution. We thoroughly explain how the current design flaws in \ac{k8s}'s \criapi could be efficiently fixed. As a stopgap solution, we propose a proof-of-concept that employs the \ac{ebpf} to intercept and mitigate rogue downloading images.

\section{Background}\label{sec:background}


This section provides the technical background for the remainder of this work. The following subsections present a small review of the \acl{k8s}, the containerized application lifecycle, and the \acl{cri}.

\subsection{Kubernetes}

\acl{k8s} is an open-source orchestration system for automating deployment, scaling, and management of containerized applications, initially developed by Google and maintained by the \ac{cncf}~\cite{kubernetes_kubernetes_2023}. \ac{k8s} is built from a set of composable modules through a set of standard \acp{api} that can be extended by the users alongside the core components~\cite{burns_borg_2016}. Over time, this allowed the development of several new infrastructure paradigms, like \ac{faas} frameworks~\cite{balla_open_2020} and multi-cluster management~\cite{carrion_kubernetes_2023}. 


\acs{k8s} uses a modular structure to implement all its services, each with a distinct function, in the cluster. The \acs{api} Server is the central communication hub for the entirety of \acs{k8s}. It processes requests, validates them, and maintains the desired cluster state. After processing the request, the \ac{api} Server is responsible for forwarding the desired state to other modules, like the Kube Controller or the Kube Scheduler. Furthermore, the \acs{api} Server enforces access controls and authentication mechanisms, ensuring secure and managed access to cluster resources.

To deploy a containerized application on \ac{k8s}, the minimum manageable object is the \pod. A \pod is a group of one or more containers that share resources, shown as a single access point for the other objects in the cluster. \pods are designed to be ephemeral, and represent a single instance of an executable resource in K8s. Rarely created as an individual object, it usually uses a more complex controller, such as StatefulSets and Deployments, to manage its deployment and replication on the cluster nodes.

\subsection{Container Runtime Interface}

\ac{k8s} relies on a well-defined \acs{api} called \acl{cri}, also called \acs{cri}-\acs{api} or just \acs{cri}, which communicates through \acp{grpc}\cite{grpc_authors_grpc_2023} with the container runtime to instantiate the \pods in the nodes. Today, the default runtime used by \ac{k8s} is \contd, also maintained by the \ac{cncf}. The Kubelet module, that is the node agent for the \ac{k8s} \ac{api}, uses the \acs{cri} to execute the operations in the container runtime. 
Finally, the container registry serves as a centralized repository for storing, organizing, and distributing container images to nodes when required. The registry additionally enables a secure and efficient way to manage the entire lifecycle of container applications. 



The \ac{cri} was developed as a replacement for Dockershim, which previously served as the bridge between \ac{k8s} and Docker. The introduction of \ac{cri} aimed to standardize communication with container runtimes, eliminating the necessity for intermediary solutions like Docker. This standardization enhances compatibility and fosters a more modular approach, allowing seamless integration with emerging container technologies such as Kata Containers and gVisor \cite{wang_performance_2022}. 

\subsection{Container Image Deployment}
\label{sec:container-image-deployment}

\begin{figure}[ht]
    \centering
    \includegraphics[width=\linewidth]{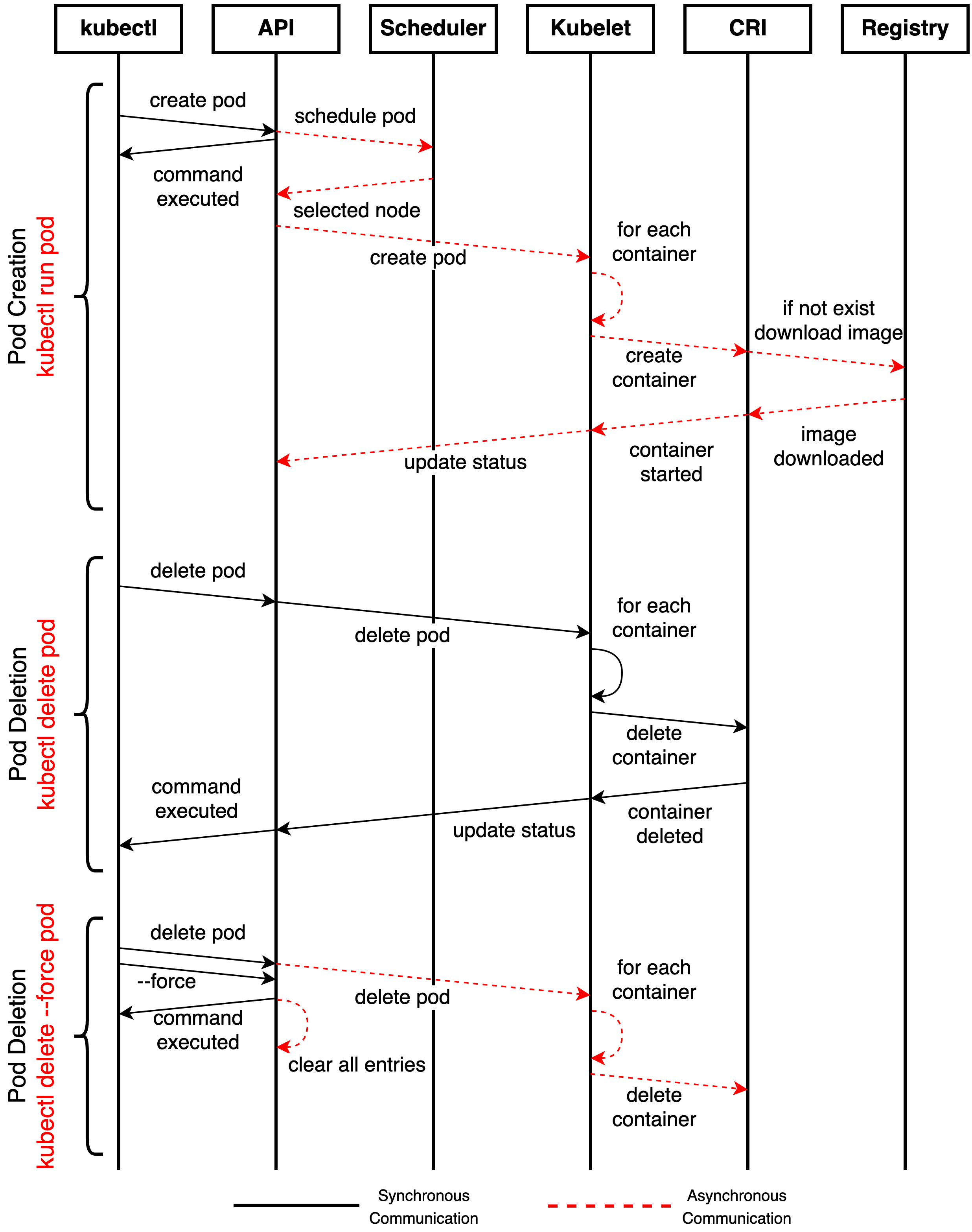}
    \caption{\pod Creation and Deletion sequence diagrams}
    \label{fig:diagram}
\end{figure}

To better understand how the \pod lifecycle is managed in the cluster, \autoref{fig:diagram} shows the sequence diagram for three \texttt{kubectl} commands: first, when a \pod deployment is initiated using the \textit{run} command; second, when that same \pod is deleted; and finally, when the deletion is forced with the \force option. In this last scenario, the \acs{api} does not wait for the \acs{cri} to confirm the deletion status. It instead immediately removes any related entries in etcd (\ac{k8s}'s database) once the command is entered.

We can break the execution of all commands into two distinct steps. Initially, the user interacts with the \ac{api} that redirects the command to the correct node. The second step is executed on the selected node asynchronously with respect to the first step. Only at the end the status of the execution is returned to the \acs{api}. While, by default, the instantiation happens asynchronously, the deletion will wait until a response is returned. To change this behavior, \texttt{kubectl} implements two options, \force and \gperiod, that execute the deletion without waiting for the \pod completion process.

Implemented to improve the agility of the cluster management, these options can, as described in the \ac{k8s} official documentation\footnote{\url{https://kubernetes.io/docs/concepts/workloads/pods/pod-lifecycle/}}, be disruptive for some workloads, or some resources may continue to run indefinitely on the cluster. We exploit this behavior to show that resources could be used by the node without any notice, mainly by how the \ac{cri} was developed. 

\section{Motivation and Attack Requirements}
\label{sec:threat}

In this section, we first present the threat model and assumptions of our work. Second, we describe our motivation. Afterwards, we describe the requirements to execute the attack on \ac{k8s}, including the configuration and image parameters that can increase the attack surface and its impact.

\subsection{Threat Model and Assumptions}

The attacker aims to increase resource usage inside a node without being noticed by the administrator. That behavior can be used as a single \ac{dos} attack or an auxiliary action to more complex threats. We assume that the attacker has access to a \ac{sa} or an user credential from a compromised Kubelet config file, or through an exploited container with a \ac{sa} linked to them. This \ac{sa} requires a minimal set of permissions for the attack: creation and deletion of \pods.

As described in the \ac{rbac} Best Practices official documentation\footnote{\url{https://kubernetes.io/docs/concepts/security/rbac-good-practices/}}, it is expected that a single cluster will have several distinct \acp{sa} to limit the access to specific namespaces, to share the cluster between production and development, or to enable a continuous integration/delivery solution. So, it is reasonable to assume that one of these service accounts could be leaked from a development machine, a non-protected staging deployment used as a vector to access an inside container with a mount \acs{sa} token, or an insider malicious user with access to the cluster. 

We also assume that quotas and limitations on the number of \ac{api} requests per second are active in the cluster. Further security configurations, e.g., image validation and locked-down private registries, can harden the attack execution but not wholly mitigate it. Indeed, a well-tuned configuration can make the threat viable even in a very constrained scenario.

On the restriction side, we understand that a compromised container or an SA token could be used for more advanced attacks, such as lateral movement \cite{minna_understanding_2021}, \ac{oom} attack \cite{larsson_impact_2020}, or deploying cryptomining in the node \cite{li_robbery_2022}. However, all these attacks can be straightforwardly identified by inspecting the resource usage or checking the elements running in the cluster. Notwithstanding, this attack wants to create a stealthy channel to cripple the cluster's resources that cannot be easily identifiable without an extensive evaluation of the node resources and processes, differentiating it from the other ones.

On the other hand, the attack can be performed with the same minimal permissions on a wide variety of settings, ranging from busy clusters with several concurrent deployments to smaller, soft multitenancy clusters\footnote{Soft-multitenancy refers to resource sharing and strong trust between the tenants. \url{https://kubernetes.io/docs/concepts/security/multi-tenancy/\#isolation}}, and multi-cluster topologies~\cite{iorio_computing_2022}. \reviewerBc{Attackers can choose to target a specific node, for example, to target a GPU-enabled node. Using a \texttt{nodeSelector} or \texttt{podAffinity}, attackers can ensure that the \pod are scheduled on the desired node(s), further increasing the attack's impact.} Even with monitoring or auditing logs active, we argue that the attack will be hard to discover without a specific and deeper analysis of the node itself, which is not a trivial task. Indeed, even in cloud deployments --- such as \ac{gke}, \ac{aks}, and \ac{eks} --- monitoring tools tend to provide an aggregated view of the cluster's resource usage, and administrators usually will not access directly the cluster nodes.

\reviewerBc{The registry employed by the attacker can be any public or private registry, as long as it is reachable from the cluster. Some public registries, like DockerHub, provide strict rate limits, which are computed per IP. Since the IP pulling the image is the node's IP, being banned translates into a denial of service for the node itself, which is exactly what the attacker wants. On the other hand, a private registry can also be used, allowing attackers to have a near-infinite arsenal of images to download.}





\generic{Finally, we} assume that the node and guest \pods have no known vulnerabilities and that all security mechanisms work properly.

\subsection{Asynchronous API Communication Status}
\label{ss:api}

As described in \autoref{sec:container-image-deployment}, \ac{k8s} uses an asynchronous communication to enforce the actions in several interfaces implemented in the cluster. For example, each \pod creation is executed independently by \ac{k8s}, but each node's \contd will have the last say in defining the order of execution for each request\footnote{The \ac{cri} specification does not mandate a specific logic for handling the order in which image pull requests are served, leading to possible differences between container runtimes.}.

This behavior can cause problems since it is necessary to control the operations' atomicity to guarantee the consistency of the cluster. Furthermore, some operations are part of a bigger dependency tree that must be executed in a certain order. For instance, deploying a container requires the download of its corresponding image to be completed (if not already available).

Since after sending the request, \ac{k8s} maintains no control over the request execution, it relies on the interface implementation to update the intermediate and final status on the object. However, this does not always happen. The \ac{cri} has no \ac{api} related to the status of image download, and so, the only response \contd will provide will be either the download confirmation or some error on the downloading process, like a missing secret or an unavailable tag.

The \ac{kep} 3542\footnote{GitHub issue available at \url{https://github.com/kubernetes/enhancements/issues/3542}} proposes an update to the Kubelet, enabling download status updates to the \ac{api}, solving the problem previously described. Even with this \ac{kep}, the problem is not entirely solved: indeed, it remains impossible to cancel pending downloads. \reviewerCc{Even using alternative container runtimes such as CRI-O or more secure ones like gVisor or Kata Containers does not change the situation. All these runtimes still rely on the \ac{cri} interface, which suffers from the issue described above.}

\section{Attack Description}
\label{sec:attack-description}

The following section proposes two attacks that exploit the aforementioned vulnerability in \contd and the \criapi, first rendering \reviewerBc{a} \acl{k8s} node unable to pull any \pod image for an extended period and, secondly, impacting a node's cache, increasing the amount of time needed to deploy an instance of an application.

\subsection{Attack Vector}
\label{sec:attack-vector}

As mentioned in \autoref{ss:api}, \acl{k8s}' \acs{api} abstracts away several steps that must be taken behind the scenes to create a \pod. When a user creates a \pod through the \acs{api} (as in \autoref{fig:diagram}), it communicates with the Kubelet to issue a \pod creation request. The Kubelet then communicates with \contd via the \ac{cri} to trigger the image pull, set up the container, and finally the \pod status is returned to the user. In normal conditions, the \acs{api} waits for \contd to respond with the status of the \pod creation before sending a response to the user. However, the \acs{api} does not wait for \contd to complete pulling the images. In fact, as soon as all required containers for a \pod have been created by \contd, the \acs{api} immediately sends a successful response to the user. 

\looseness=-1
While this choice has its roots in the separation of concerns between \acl{k8s} and \contd, it has a subtle side effect. Without having to notify the \ac{api} about the status of the image pull, \contd is completely unaware of the addition or removal of \pods through the \acl{k8s} \acs{api}. With that, a malicious actor can try to use a significant amount of resources by repeatedly instantiating new \pods in the node.
\reviewerAt{Discussion on Quotas moved from here to the end of the section.}

\reviewerAc{To further strengthen the attack, attackers can use the \force flag when deleting a \pod. As discussed in \autoref{sec:container-image-deployment}, the \force flag automatically removes all \ac{api} bindings for the resources, which means that the \pod will be deleted immediately from the cluster. On the other hand, it will not interrupt any image download that has already been started by \contd.\footnote{At the time of writing, a bug exists in \ac{k8s} that additionally prevents the deletion of associated containers when a \pod is force deleted. See \url{https://github.com/kubernetes/kubernetes/issues/119276}. The bug is not planned to be fixed.}.}

\reviewerAc{Thus, after deploying a \pod an attacker can force delete it shortly after, and avoiding a lengthy wait for the image downloads to complete.} This process can be repeated a potentially infinite number of times, in quick succession. By doing so, an attacker can clutter the \contd's download queue with useless image pull requests as long as they have image tags available to pull. Since \contd is agnostic to the \pods' existence and \acl{k8s} is unaware about what images are being downloaded, the status quo can remain indefinitely. As such, the target container runtime will continuously be forced to keep track of and download useless container images on behalf of the attacker.



\subsection{Attack Requirements}\reviewerAt{Changed from subsubsection to subsection}
\label{sec:attack-requirements}

The attack requires minimal prerequisites to be executed. First, it needs access to a cluster with standard user privileges and the capacity to create and delete at least one \pod. \reviewerAc{Any limitations imposed by the \ac{rbac} configuration or \ac{pss} are irrelevant to its execution. Indeed, as the vulnerability sits in the \ac{cri} interface,  any hardening or security measures applied to abstraction layers above the \ac{cri} will not prevent the attack.

Quotas are also easily bypassed: the minimum requirement is to be able to create and delete a single \pod in a single namespace. \ac{k8s}' Quotas are designed to limit the number of objects that can be created in a namespace by a single user, preventing resource exhaustion attacks and ensuring fair resource distribution among users. However, since the attack relies on the constant deletion and recreation of \pods, a Quota of just one \pod is sufficient to execute the attack.}


Second, the attacker requires a set of Docker images, which can be either different tags of a single image or entirely different ones altogether. They also need to be in a registry accessible by the node. Furthermore, the least layers the images have in common, the better, since re-occurring layers will not be re-downloaded.

The content, structure, functionality, and size of the image are irrelevant for the attack to be successful. Their only limiting aspect is that the total size of the image cannot exceed eventual limits imposed by the cluster administrator because the download will be blocked before starting. \reviewerDc{Thus, attackers can use any image they have access to, including public images from Docker Hub or private images from their own registries, as long as they are accessible by the node.} \reviewerDc{Such images either need not to be present on the target node, or their download must be forced using the \texttt{imagePullPolicy: always} parameter.}

On the other hand, the amount of images available to the attacker is critical for the attack's success. Since the image (or any of its layers) must not be present on the attacked node, the more are available, the longer the attack will last. This must also scale up with the amount of storage available on the node.


\subsection{Denial of Resources}
\label{sec:denial-of-resources}


This attack will generate a denial of resource in the node, and its impact is twofold. First, the endless cycle of image pull requests and deletions prevents the deployment of new containers without a cached image. This problem is exacerbated by the growing trend of running short-lived microservices: indeed, 72\% of containers live fewer than five minutes \cite{sysdig_cloudnative_2023}. An unresponsive cluster can be extremely problematic, even if the attack is short-lived: clusters that rely on short and fast batch jobs (e.g. \ac{cicd}) will inevitably be delayed. \reviewerBc{Indeed, this \ac{dos} does not render a node actually \textit{unavailable} for \ac{k8s}: the scheduler will still see the node as healthy and will continue to schedule \pods on it, nor any Pod on it will be evicted.} 

Second, the attack drains the available resources on the worker node by forcing it to constantly utilize disk, bandwidth, and CPU to retrieve container images. As a result, other workloads hosted on the node may suffer a performance setback, possibly violating \acp{sla}. Since the Kubelet usually runs as a process in the node and it is not directly controlled by the quotas and limitations imposed by the \acl{k8s}, it will use 100\% of the resources available to download and decompress the images, and by default, there is no way to limit this resource access. \reviewerBc{Furthermore, pinpointing the resource starvation to the attack is not straightforward, as the \contd process will appear as a busy process in the node, and the \ac{api} will not provide any information about the ongoing image downloads.

Ultimately, the effectiveness of the denial of service will depend on the amount of resources available in the node and the amount of nodes in the cluster. Generally, \ac{k8s} worker nodes have a limited size, and the \ac{k8s} philosophy promotes smaller but more numerous nodes, enabling horizontal autoscaling and a better failure tolerance. Additionally, \ac{k8s} is increasingly deployed in Edge environments, where nodes are often resource-constrained and have no redundancy. In these scenarios, the attack can be particularly effective, as it can quickly exhaust the available resources. On the other hand, larger nodes with more resources can withstand a higher number of image pull requests, but the attack can still cause a noticeable performance degradation.

Nonetheless, the amount of nodes in the cluster also has a direct impact on the attack. When attackers target a single or a few nodes, it can be easily attributed to a local malfunction, and its resource usage will be hard to pinpoint in an aggregated dashboard. On the other hand, executing the attack on many or all nodes in big clusters will likely cause a noticeable increase in resource usage across the cluster, making it easier to identify.}

\reviewerDc{
Finally, the possibility of a \ac{dos} is increased in multi-tenant scenarios such as Liqo~\cite{liqo_contributors_liqo_2019}, where resources are shared across clusters. As an example, a cluster may lend a subset of its resources to another cluster, allowing the latter to run workloads on it. To the renting cluster, this shows up as a local, virtual namespace, while \pods are actually running remotely. In this case, an attacker can target the shared namespace, causing a denial of service on the remote node(s).
}

\subsection{Cache Manipulation}
\label{sec:cache-manipulation}

Since the images are not deleted after the download, they can be used to manipulate the cache in a given node. The \ac{gc}'s default configuration sets a series of thresholds, the most important ones being \texttt{ImageGCHighThresholdPercent}, used to determine when the \ac{gc} will be triggered, and  \texttt{ImageGCLowThresholdPercent}, the minimum target percentage that the \ac{gc} wants to achieve after starting to delete images. The default settings for these two thresholds are 85\% and 80\%, respectively.

After surpassing the upper threshold, the Kubelet marks images present on disk for more than two minutes (by default) as eligible for deletion, and starts removing them until at least the lower threshold is reached. It must be noted that the \ac{gc} may keep deleting eligible images even after surpassing the lower threshold. This can be a problem in scenarios where the cluster hosts cron jobs, \ac{faas} tasks or any application that does not have a container continuously running. Since their corresponding images will be deleted, such tasks will suffer from a delayed execution, as \contd will be forced to download the image again.


However, \ac{k8s} does not just limit to cleaning up unused images and containers. Another key part of the system is the Node-pressure \pod eviction system\footnote{\url{https://kubernetes.io/docs/concepts/scheduling-eviction/node-pressure-eviction/}}, which is independent from the \ac{gc}. When a node's resources reach dangerously low thresholds, the Kubelet intentionally terminates \pods to avoid possible contention and crashes. At the time of writing, hard eviction is handled by the \texttt{DefaultEvictionHard} option, whose default value is 90\%. Once the threshold is reached, it will delete both \pods and their corresponding images, taking care in deleting as few of them as possible. This is meant to be an extreme measure, and usually causes further disruption on the node due to the critical lack of disk space and increased CPU usage.



\subsection{Responsible Disclosure}

\reviewerAc{The attack presented in this work has been responsibly disclosed to the \acl{k8s} community. We opened a public issue to discuss the attack and its implications\footnote{GitHub issue available at \url{https://github.com/kubernetes/kubernetes/issues/122905}}. The issue was acknowledged by the \acl{k8s} maintainers, but as of the time of writing, a solution has not yet been implemented in the codebase. Still, the maintainers have confirmed that the attack is not a security vulnerability, but rather a resource usage bug, spurring a discussion and several pull requests on how to address it in future versions of \acl{k8s}.} 
\section{Experiments and Results}
\label{sec:attack-experiments}

In this section, we present a summary of our findings after running a series of experiments. We evaluate the effectiveness of the attack on a real \ac{k8s} cluster and additionally analyze the effect of the attack on some key system resources.

\subsection{Experimental Setup}
\label{sec:cluster-setup}

\reviewerDc{To evaluate the attack, we conducted a series of experiments on a \ac{k8s} cluster. The goal was to assess the attack's impact on resource usage, scheduling delays, and service interference. We also wanted to verify the attack's effectiveness in manipulating the local image cache, and finally, to confirm the attack's applicability in public cloud environments.}

\subsubsection{Local testbed}
\label{sec:local-testbed}

\reviewerDc{
We set up a local testbed on our premises to evaluate the attack. On a machine equipped with an Intel Xeon Silver 4112 CPU (2.60GHz) and 64 GB of RAM, we created three virtualized nodes with 2 vCPUs, 4 GB of memory, 120 GB of storage capacity (HDD), a 1 Gbit/s network interface, and Ubuntu 22.04 LTS as the operating system, to run as nodes in a \ac{k8s} cluster: one for the control plane and two as worker nodes, using \ac{k8s} v1.27 and \texttt{containerd} v1.6.8 as the container runtime. We also used a separate physical machine hosting a private Docker registry.




}
\subsubsection{Public cloud testbed}

\reviewerBc{To further confirm our findings in bigger and more realistic scenarios, we also conducted some experiments in a public cloud environment. We used Google Cloud Platform's \ac{gke} service, allowing us to assess the attack's impact on a larger scale and in a more production-like environment. We used \texttt{e2-standard-2} machines with 2 dedicated AMD Epyc vCPUs, 8 GB of RAM, a 60 GB balanced persistent disk, and a 10 Gbit/s network interface. We deployed a total of ten worker nodes, while the control plane was provided by \ac{gke}. The nodes were configured to use Ubuntu 24.04 LTS as the operating system and \texttt{containerd} as the container runtime. No particular settings were enabled during the creation of the cluster.}

\subsubsection{Image sets}
\label{sec:images}

For the effectiveness and consistency of the experiment, we decided to build our own set of Docker images to be used in the attack.


Each image in the set contains a base Ubuntu layer (of around 80 MB uncompressed), which is shared between all the images, and a variable number of layers, created with the \texttt{od} command and the \texttt{c} option. \generic{We performed some experiments on \texttt{gzip}'s compression options, and the results are available in a separate report~\cite{dias_knob_exploiting_2023}.}

The first set, called \textbf{Variable GB}, contains a total of seven images. Each of the images contains one to seven additional layers of 2GB each. As a result, each image in the set ranges in size between 2GB and 14GB. The second set of images called \textbf{Variable MB}, contains forty images, with each additional layer containing 20 MB of data. This results in images ranging between 100 MB and 900 MB. These sizes refer to on-disk usage, as when fetched from a registry, they will be compressed, decreasing their size by roughly half. \autoref{tab:images} shows a summary of these two sets.

\begin{table}[t!]
    \centering
    \begin{tabular}{lcc}
    \hline
    Image set & \textbf{Variable GB} & \textbf{Variable MB} \\ \hline
    Tag count & 7 & 40 \\ \hline
    \makecell[l]{Lower-bound\\uncompressed size} & 2.12GB & 0.10GB \\ \hline
    \makecell[l]{Upper-bound\\uncompressed size} & 14.40GB & 0.90GB \\ \hline
    \makecell[l]{Total\\uncompressed size*} & 57.12GB & 16.80GB \\ \hline
    \makecell[l]{Total\\compressed size*} & 28.88GB & 8.47GB \\ 
    \hline \\
    \end{tabular}
    \caption{Summary of layers, size, and other characteristics of the two image sets used in the experiments. *The total size includes the base image only once, as it is not duplicated when generating an image. The compressed size takes into account \texttt{pigz}'s compression factor of 0.504.}
    \label{tab:images}
\end{table}

\subsection{Denial of Resources and Scheduling Delay}
\label{sec:attack-scenarios}

\begin{figure*}[t!]
    \begin{subfigure}[b]{0.47\textwidth}
        \includegraphics[width=\textwidth]{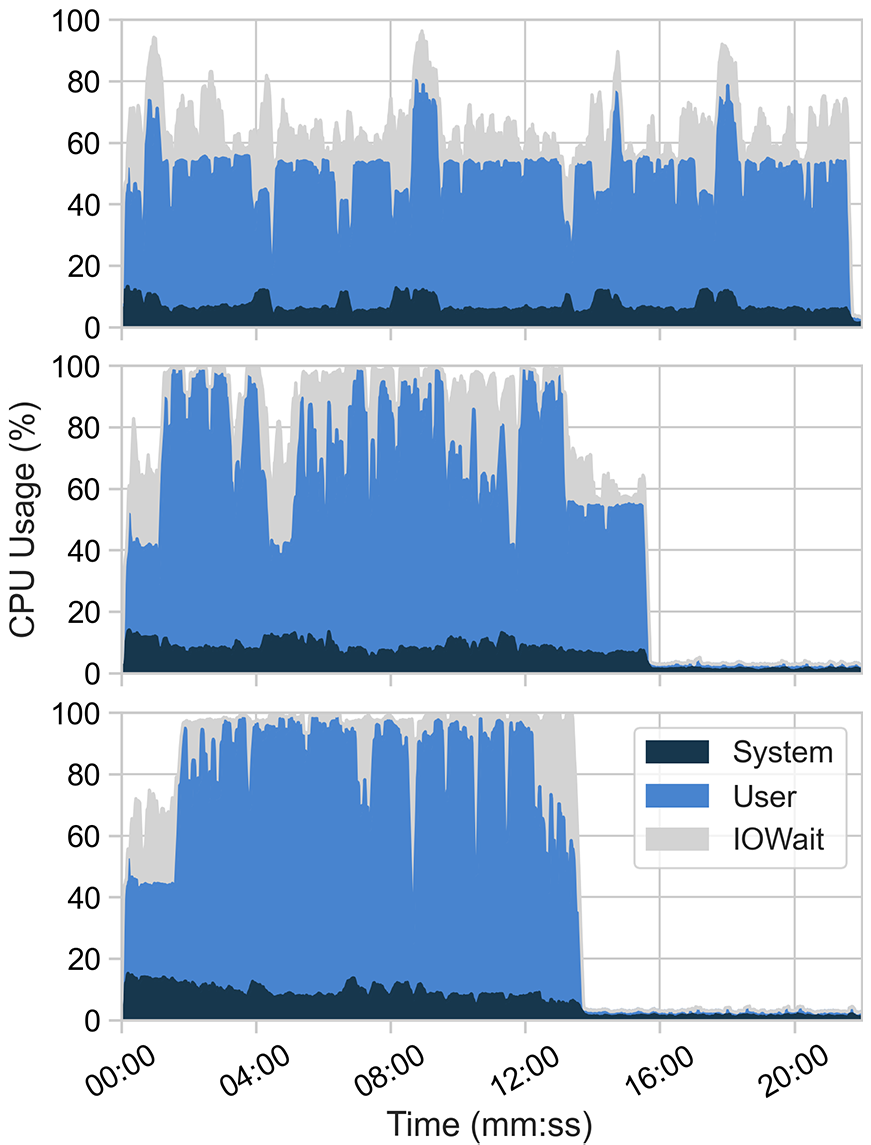}
        \caption{CPU usage by type with \mp set to 1 (top), to 2 (middle) and to 4 (bottom).} 
        \label{fig:randomgb-cpu}
    \end{subfigure}\qquad
    \begin{subfigure}[b]{0.47\textwidth}
        \includegraphics[width=\textwidth]{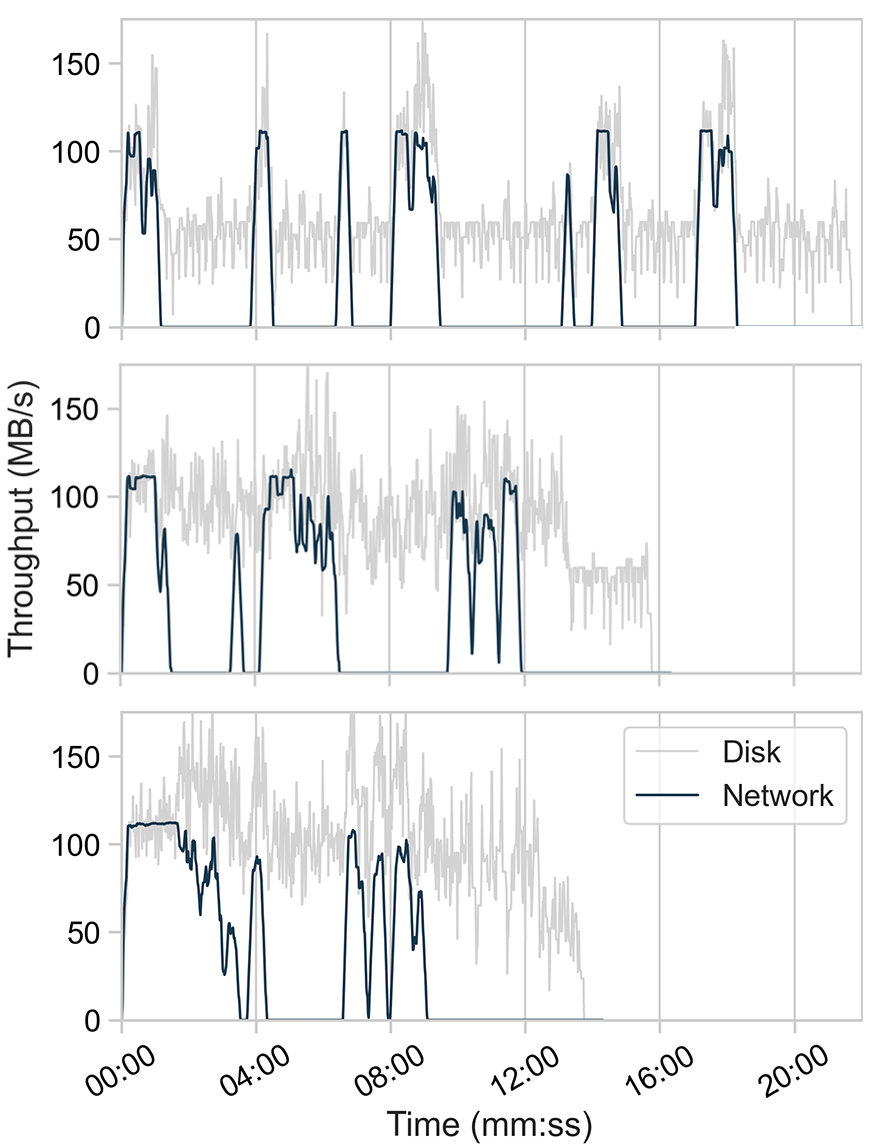}
        \caption{Disk write and network read with \mp set to 1 (top), to 2 (middle) and to 4 (bottom).}
        \label{fig:randomgb-disknet}
    \end{subfigure}
    \caption{CPU usage by type (on the left); disk write and network read (on the right) of the system during the variable GB scenario. Data were taken from a randomly sampled trial.}
  \end{figure*}

\generic{We first focused} on assessing the denial of resources capabilities of the attack. In particular, we investigate\generic{d} to what extent the \pod scheduling can be delayed and how the CPU and I/O are affected. \generic{These tests were performed on the local testbed.}

\subsubsection{Scheduling Delay Metric}
\label{sec:scheduling-delay}

To properly compare the different lengths of the attack and the actual impact on the node image pulling, we created a metric that would normalize the results. We call it \textbf{Scheduling Delay}. It is measured in s/GB and is calculated as follows:

\[
    SD = \frac{D}{S_{c}}
\]

Where $D$ is the duration of the attack (in seconds), and $S_{c}$ is the total amount of compressed data that the node had to withstand from the beginning of the attack. This metric is unrelated to the impact on system resources and gives us a rough idea of how efficient is a certain attack configuration in preventing a node's downloading of new \pod images.

\subsubsection{Max Parallel Image Pulls}
\label{sec:max-parallel-image-pulls}

Starting from version \texttt{v1.27}, \ac{k8s} allows users to specify the maximum number of images to be pulled in parallel\footnote{GitHub issue available at \url{https://github.com/kubernetes/enhancements/issues/3673}}; indeed, in the past, images were exclusively pulled serially. While this feature helps improve the efficiency of the cluster by making full use of the available bandwidth, we want to investigate its effects on the attack since it would prevent the cluster from being stuck in a given download. However, this option would probably increase resource consumption.

In the experiment, we use the values 1, 2, and 4 of the \maximgp flag. This flag will be shortened to \mp in the upcoming sections for brevity.

\subsubsection{Results}
\label{sec:attack-results}

\textbf{Variable GB set.} We first run the experiments of the Variable GB image set \reviewerBc{on the local testbed}. Before each experiment, we reshuffle the images. Then, for each image, we deploy it, wait two seconds, and then delete it with the \force flag. Additionally, to verify the attack's behavior with the parallelism provided by the \mp flag, we run each experiment with the values described before. We repeat this process thirty times for each setting; the final result is an average of these runs. To better visualize the results, we chose a random trial from the thirty and plotted the CPU usage, broken down by type, and the disk write and network read. These plots are available in \autoref{fig:randomgb-cpu} and \autoref{fig:randomgb-disknet}. 

With \mp set to 1, the results are staggering. Even with a small attack, no parallelism and just seven images, \reviewerBc{the impact on both worker nodes is substantial, and clearly could be categorized as a \ac{dos}.} Every new download on the node is blocked for a total of 1352 seconds, almost 23 minutes; the CPU usage averages 65\% with spikes of almost 90\%, and the disk is kept constantly busy, with I/O write spikes of 150 MB/s. From \autoref{fig:randomgb-cpu} (row 1), we can see how these CPU peaks are noticeable and correlate with the download of the images (\autoref{fig:randomgb-disknet}), and thus with a high bandwidth and I/O usage. On the other hand, when images are decompressed, the CPU usage drops significantly but remains steadier, as does the disk usage.

As we increase \mp, the total time of the attack decreases, and the resource usage increases accordingly. For example, setting \mp to $2$ reduces the attack time to 909 seconds, while with \mp$=4$, it further reduces to 887 seconds. On the other hand, the CPU usage increases, while the difference in the peaks becomes less and less apparent as the various parallel downloads compete for CPU and I/O time with the ongoing unpacking tasks. With four parallel images, the CPU usage reaches almost 100\% for most of the attack time.

By calculating the Scheduling Delay for these three tests, we see how it also drops from 46.82 (\mp$=1$) to 31.48 (\mp$=2$) and 30.72 (\mp$=4$) s/GB. It is important to observe the difference between the former two settings, where the SD drops 33\%, and the latter, where the SD drops again but by a smaller margin (only 2.4\%). While introducing parallelism strengthens the attack and puts an increased load on resources, at a certain point, the returns start diminishing as contention is introduced, and the cluster's overall performance starts degrading. Thus, we believe that increasing \mp without implementing a solution to this threat could be even more detrimental.

\textbf{Variable MB set.} We now repeat the same experiments on the Variable MB set. Such results are similar to the ones we obtained with the Variable GB set, but with a much lower runtime due to the overall smaller size. Thus, we calculate the Scheduling Delay metric, presented in \autoref{tab:scheduling-delay}, and the average CPU usage over all cores, presented in \autoref{tab:cpu-usage}, to properly assess the consequences of using smaller-sized images in the attack. 

\begin{table}[t!]
    \centering
    \begin{tabular}{lccc}
    \hline
    \multirow{2}{*}{Set} & \multicolumn{3}{c}{Scheduling Delay} \\
                     & \texttt{mp}$=1$     & \texttt{mp}$=2$    & \texttt{mp}$=4$ \\
    \hline
    Var. GB & $46.82 \pm 0.37$ & $31.48 \pm 1.02$ & $30.72 \pm 0.72$ \\
    Var. MB & $56.93 \pm 1.59$ & $41.10 \pm 1.16$ & $39.92 \pm 1.60$ \\
    \hline \\
    \end{tabular}
    \caption{Scheduling Delay (s/GB) averaged over 30 trials with standard deviation. The higher, the more disruptive the attack.}
    \label{tab:scheduling-delay}
\end{table}

\begin{table}[t!]
    \centering
    \begin{tabular}{lccc}
    \hline
    \multirow{2}{*}{Set} & \multicolumn{3}{c}{CPU Usage} \\
                     & \texttt{mp}$=1$     & \texttt{mp}$=2$    & \texttt{mp}$=4$ \\
    \hline
    Var. GB & $67.31 \pm 0.81$  & $91.67 \pm 2.24$ & $93.22 \pm 2.23$ \\
    Var. MB & $64.14 \pm 0.35$ & $84.35 \pm 0.62$ & $85.78 \pm 2.69$ \\
    \hline \\
    \end{tabular}
    \caption{Average CPU usage over all cores of a worker node during the attack for each scenario. Factor of 100, averaged over 30 trials.}
    \label{tab:cpu-usage}
\end{table}

From the Scheduling Delay results, the immediate impression is that using smaller images is indeed more disruptive. With \mp$=1$, the Variable GB set hangs the \reviewerBc{worker nodes} for 47 seconds per GB. On the other hand, the Variable MB manages to block it for 57 seconds. Thus, we can block the downloads for 18\% more time with small images. With \mp$=2$ and \mp$=4$, this becomes more evident, arriving at 23\% more with small images. We attribute this increase to the overall overhead that \contd must withstand when downloading images. Indeed, each new image implies fetching its manifest, obtaining the list of layers, and manually downloading each of them: all operations repeated many times start piling up.

On the other hand, \autoref{tab:cpu-usage} presents almost opposite results. While with no parallelism, the difference is almost unnoticeable, as \mp is increased, the Variable GB set is far more CPU-intensive, using almost 10\% more CPU on average than the Variable MB set. We attribute this lower average CPU usage to the relative weight of downloading and unpacking a smaller image. As a consequence, the possibility of saturating the CPU is lower unless the parallelism is way higher than in our experiments.

\subsection{Service Interference}
\label{sec:service-interference}

\begin{table*}[t!]
    \centering
    \begin{tabular}{lcccc}
    \hline
    \multirow{2}{*}{Benchmark} & \multicolumn{2}{c}{No attack} & \multicolumn{2}{c}{With the attack} \\
                           & Results        & Trials       & Results           & Trials \\
    \hline
    \texttt{kernel} & $823.25 \pm 0.41 \: s$  & 3 & $1630.04 \pm 5.47  \: s $ & 5 \\
    \texttt{stress} & $2185.83 \pm 1.29 \: bo/s$ & 3 & $1463.18 \pm 12.52 \: bo/s $ & 15 \\
    \hline 
    \end{tabular}
    \label{tab:benchmark-results}
    \caption{Summary of the results of the benchmark on the non-disrupted and on the attacked cluster. $bo/s$ stands for bogo-operations/seconds.}
\end{table*}


Another goal of the attack is to affect the \ac{qos} of other running services in the node. We perform\generic{ed} two experiments \generic{on the local testbed} with the same attack, but this time, we verif\generic{ied} how this affects CPU and I/O usage of other \pods.

To do so, we employ the Phoronix Test Suite\cite{phoronix_media_phoronix_2023}, an automated benchmark suite that allows users to perform a wide range of performance assessments on their system. From the wide variety of benchmarks available, we choose the \texttt{build-linux-kernel}\footnote{Specifications available at \url{https://openbenchmarking.org/test/pts/build-linux-kernel-1.15.0}. The ``defconfig'' option was used.} and \texttt{stress-ng}\footnote{Specifications available at \url{https://openbenchmarking.org/test/pts/stress-ng}. The  ``cpustress'' option was used.} benchmarks, two of the most popular of the suite. The first assesses the time needed to compile the Linux kernel (version 6.1) on the machine, putting an I/O and CPU workload on the system. The latter evaluates the amount of operations per second the CPU can perform. To ensure the statistical relevance of its results, Phoronix automatically performs additional trials if it encounters performance issues or suspiciously high variability between different runs.  


First, we run the \texttt{build-linux-kernel} benchmark with no attack running on the cluster. In this case, it completes in 853.25 seconds, averaged over three runs. Then, we apply the attack as described in \autoref{sec:attack-scenarios} with the Variable GB set and \mp set to 1 (as parallelism is not a concern in this setting). We chose this particular scenario as it is the longest and, thus, the most likely to last the entirety of the benchmark. We therefore re-run it, applying the attack in the background and cleaning up after every trial. With this setup, the build time increases to 1630.04, a 91\% increase over the non-disrupted one.

The other benchmark also reports significant results, with the \texttt{stress-ng} measuring 2185.83 bogo-ops/s\footnote{Bogus operations per second, a measure used by \texttt{stress-ng} to evaluate processor performance} with no attack running over three trials. Repeating it with the attack, Phoronix takes 15 trials to obtain 1463.18 bogo-ops/s, a drop of 33\% with regards to the non-attacked scenario. Additionally, the standard error is 12.52 seconds, almost ten times higher than the 1.29 seconds of the non-attacked benchmark. This 33\% drop is less severe than the one from \texttt{build-linux-kernel}, mainly due to the fact that \texttt{stress-ng}'s CPU-bound behavior clashes more with the attack, introducing additional contention.


\subsection{Cache Manipulation}
\label{sec:cache-manipulation-exp}

\begin{figure}[t!]
    \centering
    \includegraphics[width=\columnwidth]{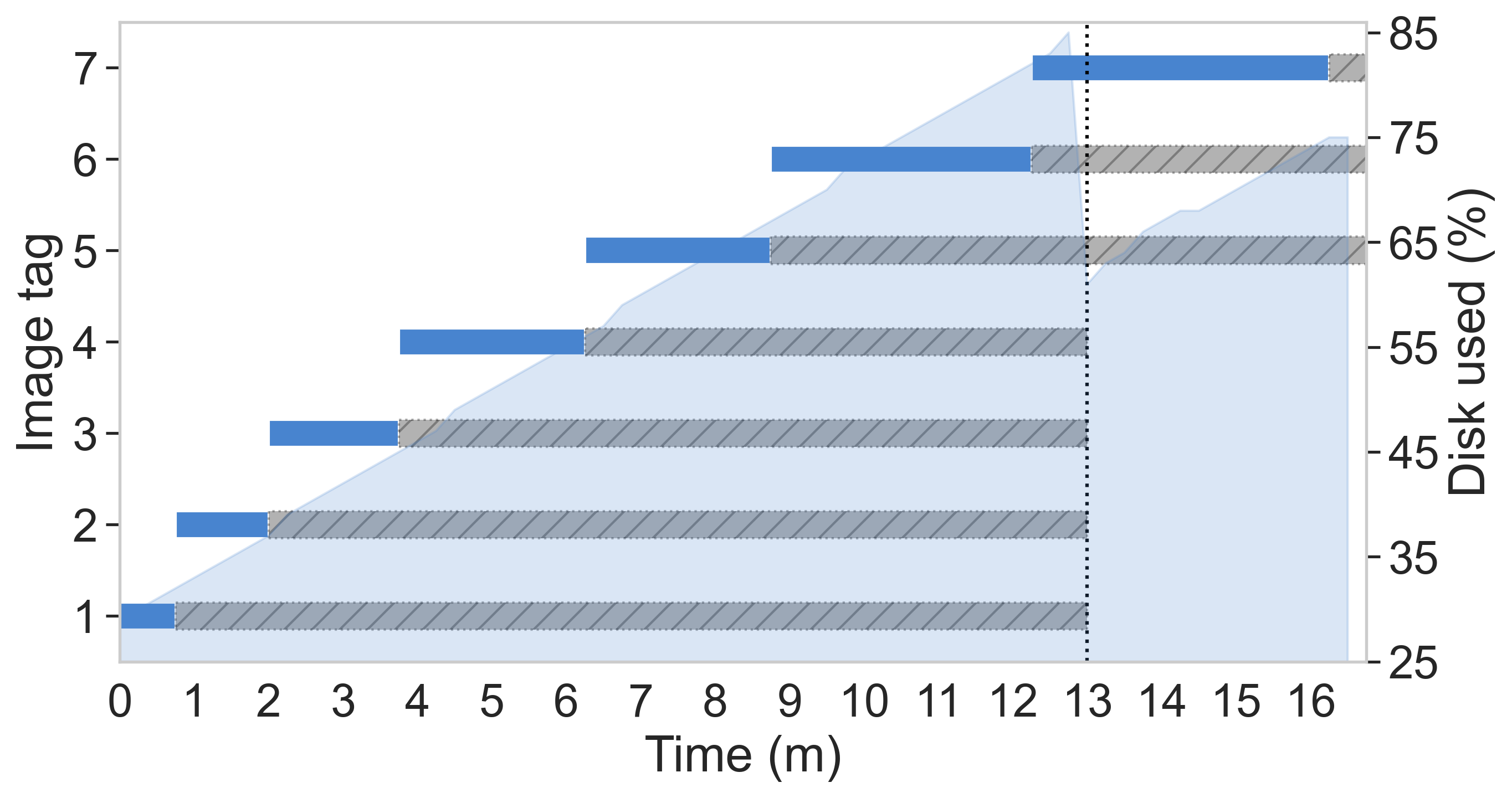}
    \caption{Sequence diagram of downloaded images (left axis) and used disk space (right axis, shaded area) during the sequential image download. Dark lines represent active downloads; grey lines represent image locality (i.e. image finished downloading and present on disk).}
    \label{fig:gc}
\end{figure}

\generic{We then verified} the attack's capabilities in interfering with another kind of system resource: storage. In particular, we analyze how the attack can manipulate the local image cache \generic{on the local testbed.}

As previously described in \autoref{sec:cache-manipulation}, \ac{k8s} is equipped with a local image cache which is periodically expunged by a \ac{gc} should the used disk space surpass the \texttt{ImageGCHighThresholdPercent} (85\%) threshold. Thus, we aim to investigate whether the attack can induce cache evictions as a side effect. Such a scenario would mean that the attacker can have a way of indirectly controlling the cache. In addition, it would further disrupt the node's operations as the I/O and CPU overhead of the cache eviction is added to the underlying attack.

We therefore rerun the Variable GB experiment with \mp set to 1, as we are not interested in seeing the effects of parallel downloads in this case. With a simple script, we monitor the images present in the cache and the disk space. However, this time, we run the attack sequentially rather than randomly to eliminate variability. In other words, first we deploy the smallest image (2 GB), delete it with \force, deploy the second smallest, and so on. We repeat the experiment several times, obtaining varying results. An example result is shown in \autoref{fig:gc}.

Indeed, we discovered that the \ac{gc} is extremely aggressive, and once the 85\% threshold has been reached (at $t=13$), it starts deleting all qualifying images that are eligible for deletion. In our case, during the download of the seventh image the \ac{gc} deleted images numbered 1 to 4 at once, dropping the disk usage to 62\%. Since the minimum \ac{ttl} of images downloaded to disk is two minutes, the sixth image was not removed. On the other hand, the fifth image was  spared for no reason, and in other trials, it was also deleted.

Either way, this greedy behavior by the \ac{gc}, while beneficial for the disk pressure, further bolsters the capabilities of the attack.  Since in the background more images are being downloaded, saved to disk and unpacked, triggering the deletion of several images at once causes further impact on system resources and potentially deletes images that may have been needed for short jobs by other tenants.

On the other hand, we obtained a slightly different result by avoiding to delete \pods at all. Doing so prevents the \ac{gc} from deleting any image --- as they are all in use --- and forces it to wait for the eviction manager to delete at least one \pod. Consequently, the disk usage eventually reaches 90\%, triggering the eviction manager. As it evicts \pods, the \ac{gc} also deletes their image shortly after. This overlap of both the eviction and \ac{gc} causes even more mayhem on the node, heavily impacting CPU and storage I/O. 

\reviewerBt{Attack Feasibility moved to Denial of Resources in Attack}

\subsection{Applicability to Public Cloud}
\label{sec:scalability}

\reviewerBc{
To further confirm our findings, we conducted additional experiments in the \ac{gke} public cloud scenario, aiming to determine if the attack was feasible in larger and more realistic scenarios. First, we replicated all the experiments described so far in this section. The results were consistent with those obtained in the local testbed, but due to space constraints, they can be found in the repository. Second, to investigate whether increasing the number of nodes affects the attack's capabilities, we crafted a new scenario using a set of real, \ac{ml}-oriented images and a slightly different attack procedure. Additionally, information about the setup can be found in the repository.

In this scenario, instead of creating and deleting \pods, we first create a Deployment. The Deployment is configured to have eight replicas, one for each node in the cluster, using affinity features to achieve so. Since the images are large, we wait several seconds after the first creation, and then we patch the Deployment, changing the image. As a result, the underlying ReplicaSet is substituted, triggering a process in which old containers are deleted, and new ones are spawned with the new image, thus requiring a new download from the container registry. This process realizes the same effect as the previous attack with the \force flag, but without actually using it. We also validate whether Falco can identify this new approach. Yet, with the default set of rules, the only \texttt{INFO} alert printed is the one related to the first Deployment creation, since ReplicaSets are not observed by Falco.

This attack proves, in a production scenario, that all the characteristics explored in the local testbed (the scheduling delay, resource usage, and cache manipulation) are still valid. With just a single alert triggered, an attacker could perpetually edit the Deployment, keeping the node(s) under constant pressure. The CPU usage over all nodes is still high, averaging 91\% over the attack duration. Each node reacted differently to the various download requests: unfortunately, some images were never downloaded by some nodes. Yet, all of them were overall delayed for a 18 minutes on average. The \ac{gc} could have also been triggered if we had increased the number of images. Thus, we can conclude that the attack is indeed feasible in a public cloud environment. Attackers wishing to make it stealthier could reduce the amount of replicas or use specific nodeSelectors, achieving a similar effect. The only required permission is the ability to create and patch Deployments, and the number of nodes that can be affected is limited by the amount of replicas one can create. 
}




\begin{figure*}[ht!]
    \centering
    \includegraphics[width=\textwidth]{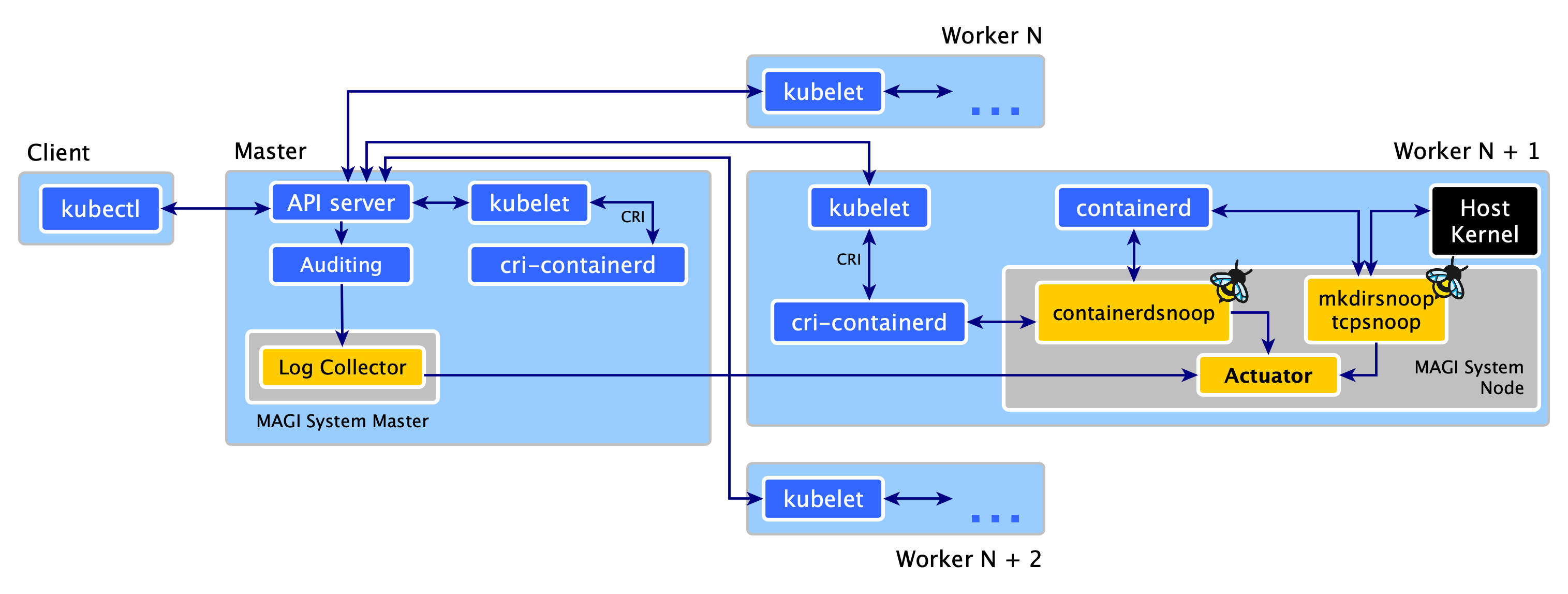}
    \caption{Architecture of the MAGI System. Dark components are part of \ac{k8s}; light components are part of MAGI. Arrows represent communications and data flow.}
    \label{fig:system-diagram}
\end{figure*}

\section{Mitigation Discussion}
\label{sec:solution}

This section presents the rationale about how the problem can be solved in the \acl{k8s}, \contd, and \criapi code, the difficulties in implementing the solution, and, as an alternative, we present a proof-of-concept solution, using eBPF until all the necessary implementation has been released inside the respective projects. 

\subsection{Kubernetes Status and Best Approach}

To some extent, the limitations of the \criapi in controlling/reporting the image downloading process are already known by the community. As mentioned in \autoref{ss:api}, KEP-3542 has already been assigned to implement an image progress notification. Open since 2019, the reasoning behind this proposed enhancement is to bring more information to the administrator, reporting the current speed and download status from each image. Nonetheless, the threat previously described will not be solved by this implementation since the progress notification will not enable the termination of the downloading process if a container has been marked for deletion. 

On the other hand, KEP-3673 outlines the Parallel Image Pulls option (outlined in \autoref{sec:max-parallel-image-pulls}) and its role in preventing cluster saturation due to prolonged image downloads. We show in our experiments how in reality this feature can actually increase the effectiveness of the attack: with more images to be downloaded concurrently, the attack will consume more resources.

As discussed in \autoref{ss:api}, the problem is the lack of communication status routes in the \criapi, because there is no status message between the user's request (create or update) and its completion. The difficulties to completely solve this issue are threefold. First, ensure that the \contd maintains the context open during all the command executions. Second, implement a series of new \ac{api} routes to guarantee that the \ac{cri} can cancel the submission or the download of each requested image to the runtime. Finally, the \pod controller must also be adapted to include these new routes, maintaining the status of the requests, validating if the image download has already started, and deciding if it needs to stop after a delete request.

Unfortunately, these implementations are not trivial and depend on several interactions with different Special Interest Groups (SIGs) and projects with distinct schedules and contribution methodologies. We are already collaborating with them to solve these issues. However, the implementation process is usually very long and laborious. Therefore, we developed a temporary mitigation solution that is presented in the following subsection.

\subsection{MAGI System}

As a stopgap and proof-of-concept solution, we propose the \textbf{MAGI System} (Mitigation AGainst Images) \cite{dias_knob_magi_2023}, whose architecture is shown in \autoref{fig:system-diagram}. \reviewerCc{MAGI is fully deterministic and is written in eBPF, Python, and Bash. It monitors the cluster status and image downloading requests, and uses kernel-level tools to terminate offending image downloads whether an attack matching our pattern is detected.}


The MAGI System is composed of two main components: the \textbf{master component} and the \textbf{node component}. The master component, as the name suggests, must run on the master node of the cluster; node components must be run on each node of the cluster. This includes the master itself, if users wish to schedule \pods on it.

\reviewerDc{MAGI can be deployed in any \ac{k8s} cluster, regardless of the container runtime being used. It requires minimal effort to be set up, with the only requirement being the enabling of \ac{k8s}' auditing feature and root access to the nodes. Once deployed, it runs in the background and does not require any user interaction.}

\subsubsection*{Master component}

The master component leverages \acl{k8s}'s auditing capabilities to keep track of the \pod creation and deletion requests, which are handled by the \acs{api} server. As a pre-requisite, thus, it requires enabling \acl{k8s}'s Auditing feature\footnote{\url{https://kubernetes.io/docs/tasks/debug/debug-cluster/audit/}}, which by default is disabled. The Auditing features lets cluster administrators specify with great flexibility what to monitor, for example user activity, \acs{api} usage, and control plane events. For our purposes, we limited the scope and verbosity of the logging to only the \texttt{pods/*} resource group. More precisely, we captured all \ac{api} calls related to the \pods group and subgroups, logging event metadata. In the case of  elaborate topologies with multiple master nodes, it is sufficient to place the master component in a node in which the auditing capability has been enabled. 


Once the Auditing is properly set up, the master component can be started. During regular usage, it will monitor the \ac{api} server's calls for \pod creations, keeping a cluster-wide track where each \pods have been scheduled and if any image pull is necessary for them. Once the image has been downloaded and the \pod has been successfully deployed, the component discards the collected information.

When the component instead detects a \pod deletion request, it checks if the \pod had previously requested an image pull which has yet to complete. Such an event could potentially signal an attack attempt. In this case, the master component promptly contacts the node in which the \pod was scheduled, informing it that the tracking image should not be downloaded anymore.

\subsubsection*{Node component}

The node component is responsible for several concurrent tasks, and works with three threads. The first is in charge of keeping track of the images being downloaded by \contd; the second monitors the \ac{api} calls being made by the Kubelet to \contd; the third listens for the alerts emitted by the master component, eventually performing corrective actions. Each instance of the node component operates independently from the others.


\textbf{Image download queue monitoring.} The first thread keeps track of \crictl's local image download queue. This information is fed to the script by \texttt{containerdsnoop}~\cite{dias_knob_containerdsnoop_2023}, a fork of \texttt{dockersnoop}~\cite{ng_dockersnoop_2023}, an eBPF program capable of intercepting \ac{grpc} calls to and from \contd; in particular, calls from the ImageService (\crictl) and the RuntimeService (Kubelet).

Within the implementation, \ac{grpc} calls are not encrypted, but the \texttt{HTTP2} \texttt{HPACK} header compression mechanism requires access to the initial packets of connections. Thus, before the startup, the Kubelet instance is deliberately restarted, allowing our tool to fully parse the calls. While the program can parse almost every \ac{api} call made to \contd, we narrowed down its scope for our tool, deliberately snooping only requests and responses to the PullImage \ac{api} route.



When the tool detects a request, the payload of the call is inspected and the requested image is appended to the queue. When a response is detected -- signaling a successful download -- the said image is removed from the queue. The tool is able to distinguish between Docker Hub and other-registries images, and maintains the ordering of the download queue even when the \maximgp parameter is set to a value greater than 1.

\textbf{System call log monitoring.} A second thread is tasked with monitoring the system calls being made by \contd. Once we know that a certain image is being downloaded, we also need to identify which connection is being used to perform the download of which layer. Unfortunately, \contd does not expose such information natively. However, we fortunately noticed that \contd's behavior is predictable.

When \contd downloads an image, it first opens a connection to the registry, retrieving the manifest file and obtaining the list of layers that compose the image. Then, for each layer, \contd spawns a thread that creates a folder on the disk for storing the downloaded files, then immediately after binds to a network socket. By default, \contd opens a maximum of four sockets at a time\footnote{This option is customizable within \contd, and is unrelated with the \maximgp \ac{k8s} flag, which instead refers to \textit{parallel images} rather than \textit{parallel layers}.} when downloading images.

To properly log this information, we used a custom-made eBPF-based tool \cite{dias_knob_imagesnoop_2023} that specifically monitors two selected system calls. In particular, we attached eBPF keyprobes to the \texttt{mkdirat} and \texttt{tcp\_connect} system calls being performed by \contd. 
The first is related to the creation of the folders containing the layer data on disk and the latter with the opening of the connections to download the data in question. Since the two steps are performed by the same \contd thread, by correlating the two calls we are able to identify the mappings between \contd threads, layers, and open ports. This effectively allows us to understand which connection is being employed to download which layer. The two tools continously fetch new data and push it to userspace using a \texttt{perf\_buffer}.

\textbf{Alert listening and actuator.} Finally, a thread is tasked with listening for the master's alerts. When such an alert arrives, the component promptly reviews its local image pull queue: if the requested image is currently being downloaded, the component terminates the connection between the node and the registry, interrupting the download and clearing the local cache.

This is done first by querying the \texttt{bbolt} database\footnote{\texttt{bbolt} is a fast on-disk database used by \contd.} and obtaining the list of layers from the name of the image. Alternatively, the component can also query the registry directly, obtaining the list of layers from the manifest file. From each layer in the list, the port number of the connection downloading that particular layer is obtained. Finally, each of the offending sockets is terminated using \texttt{ss -K}. Since the Kubernetes \ac{api} has already deleted the \pod and freed up its resources, it will not attempt again to download the image, thwarting the malicious actor's attack. Clearly, if the attacker were to try again to download said image (or any other one), the system would get triggered again, thus being able to continously take down attack attempts over time.

To account for pending downloads, we put offending images signalled by the master component in a blacklist. This behavior is triggered when an image download is requested but one or more images are in the queue before it. As the queue empties, the script automatically detects if a new download matches an image in the blacklist and terminates it accordingly.

\subsection{Evaluation}
\label{sec:magi-evaluation}

To evaluate the effectiveness of our mitigation approach, we devised a series of tests that evaluate the functionality of the system in different scenarios: without the attack, with the attack, and with the attack and the mitigation running. 

\looseness=-1
Each scenario comprises the deployment of a series of images in a \ac{k8s} cluster at pre-defined intervals. We selected some common images with different sizes, to represent a common cluster scenario. While in real clusters workload might be scheduled more irregularly and unpredictably, we devised a series of ordered commands with a constant wait time between each other, allowing us to accurately assess the behavior of the cluster and our solution while minimizing external interferences. Since the \mp parameter is not the focus of this experiment, we set it to 1.

In particular, in each scenario we perform the following steps. First, to emulate the deployment of a large-sized image, we deploy \texttt{sagemathinc/cocalc} in background and wait 20 seconds to give \ac{k8s} sufficient time to complete the deployment and propagate the changes. Then, depending on the setting, we either start the attack (using the variable GB image set) or do nothing, and we wait 20 further seconds. Finally, to simulate the download of small images, we deploy \texttt{jupyter/scipy-notebook}, \texttt{nginx}, and \texttt{online-boutique}~\cite{google_cloud_platform_microservices_2023}, waiting 20 seconds between each other. In particular, \texttt{online-boutique} is a collection of twelve lightweight microservices deployed as separate containers.

We repeat these steps for each of the aforementioned scenarios. In each one, we measure the total length of the test, the moment at which every deployment was requested and the moment at which the container was actually started. This information is collected using the in-built \ac{k8s} \acp{api}.



In the first scenario, the total time spent by the test is around eleven minutes. Of these, eight of them are spent downloading the \texttt{cocalc} image, which is extremely large; once the download is completed, the other images follow suit quickly. On the other hand, in the second scenario we deploy the attack after the first twenty seconds, thus queuing it after the \texttt{cocalc} download. Once \texttt{cocalc} finishes downloading, all the following requests are brought to a grinding halt as the attack starts \reviewerBc{saturating the CPU and bandwidth of the worker nodes.} This increases the length of the test to around twenty-nine minutes. It is important to note that the attack does not affect the net download time of the images; rather, it makes them stall and wait until their turn arrives. Finally, in the third scenario, the MAGI System successfully intercepts and blocks six out of seven images, i.e., all images but the smallest one. As the downloads get interrupted almost instantly, the total length of the test drops back to eleven minutes, the same as the first scenario.

These tests \reviewerDc{highlight the capabilities and low impact of our solution.} First, we observed how MAGI has almost no footprint. The CPU usage in nodes spikes at 0.8\% during detection and almost 0\% when sitting idle. The memory footprint was more evident, ranging from 380 MB idle to almost 420 MB during the attack, but remaining mostly stable nevertheless. We attribute this usage to Python's memory management and the need of storing status for several different metrics at once. On the other hand, both CPU and memory usage were insignificant on the master node, with the only discernible overhead being \ac{k8s}' auditing feature.

\reviewerCc{Wishing to further investigate why MAGI failed to block the smallest image of the attack, we performed a series of tests with a set of smaller images, starting from 83 MB (42.5 MB uncompressed) and adding a new 5 MB layer at each step. With MAGI running in the background, we deployed and deleted a \pod with the image, waited for the download to complete, purged the node's image cache, and then repeated the process with the next image. We observed that the cutoff in which MAGI was able to block the download was around 250 MB, which on a 10 Gbit/s network takes around 2 seconds to download. MAGI's eBPF code identified all image downloads and the corresponding layers, but with such a large bandwidth, downloads complete so quickly that is almost impossible to terminate the socket before the download finishes. With smaller bandwidths, the download time increases, shifting the size threshold accordingly.
}

\reviewerBc{
\subsection{Discussion and limitations}

While MAGI demonstrates a significant improvement over the current state of affairs, it is does not come without drawbacks. First and foremost, the attack surface of the cluster increases significantly when MAGI is deployed. This is due to several factors.

First, the \ac{k8s} auditing feature -- which is not enabled by default -- exposes sensitive information about the cluster's state and operations. Second, MAGI needs to monitor the gRPC calls between the \ac{api} server and the container runtime. Doing so requires restarting the Kubelet, which can disrupt the cluster's operations. Finally, the MAGI client requires root access to the nodes in order to intercept the system calls and kill the offending sockets. All of these extra requirements increase the attack surface.

In addition to the security risks, MAGI also has some limitations in terms of performance and usability. As shown above, MAGI struggles in blocking the download of very small images. Indeed, the time required to perform heavy tasks such as the domain resolution, manifest retrieval, and socket termination can exceed the time required to download the image itself, if it is small. MAGI is also inherently fragile, relying on code and requirements that may break between Linux, \ac{k8s}, and \contd versions.} In our opinion, the sheer amount of information required for gracefully terminating a download without disrupting any service, the inherent unpredictability of the system scheduler and the relative opacity of \contd's download management render any further attempts in increasing MAGI's speed better spent in improving the main problem instead.

\section{Related Work}
\label{sec:related}

\textbf{\acl{dos} based on Resources Usage.} Resource-based \ac{dos} in Cloud Computing was previously covered in previous works. In \cite{varadarajan_resource-freeing_2012}, the authors present a novel class of attacks called Resource-Freeing Attacks (RFAs) and create a methodology to prevent the \ac{dos} by increasing the amount of resources available to the victim VM. \reviewerDc{Fang et al. also discussed resource and co-location attacks in \cite{fang_heteroscore_2023} and \cite{fang_repttack_2022}, where they proposed respectively one new metric, called Heteroscore, and a threat method, REPTTACK. In the latter, a mitigation technique is also proposed, based on introducing randomness in the scheduling of the \ac{vm}s. However, both works do not consider container-based deployments, such as \ac{k8s}.} Finally, in \cite{zhan_coda_2023}, Zhan et al. show a framework that aims to detect CPU-exhaustion \ac{dos} attacks in containers called Coda. However, it does not consider the node resources, just the container ones. \reviewerDc{Finally, some CVEs related to resource-based \ac{dos} attacks in \ac{k8s} have been reported, such as CVE-2022-1708\footnote{CVE-2022-1708: \url{https://nvd.nist.gov/vuln/detail/cve-2022-1708}}, and CVE-2025-0426\footnote{CVE-2025-0426: \url{https://nvd.nist.gov/vuln/detail/cve-2025-0426}}. Both CVEs deal with node resource exhaustion, the first one exploiting a \ac{cri} vulnerability causing large files to be written to disk, the second exploiting unauthenticated requests to cause a \ac{dos} condition in the node.}

\textbf{\acl{k8s} and Container Security.} Ahmed \cite{ahmed_docker-pi_2020} shows that the container deployment process can generate significant resource usage, mainly memory and CPU. The author also tested several alterations to the container deployment's decompress process, focusing on the parallelism of several simultaneous layers. Such approach decreases the total time to deploy a given image but inherently increases the resources used in this process. 

\reviewerAc{Additionally, two studies investigate vulnerabilities in the configuration of various services within the \acl{k8s} orchestrator. He and Guo \cite{he_cross_2023} show a threat in the resource configuration, where using highly privileged configurations allows eBPF applications to escape containers and access sensible information. Xiao \cite{xiao_attacks_2023}, instead, found several threats based on the communication between microVMs solutions --- such as Firecracker and Kata Containers --- and \contd to escape attacks and resource-based \ac{dos}.}

Another important area that has received recent attention from the container orchestration community is anomaly detection in cluster resource usage. Almaraz-Rivera \cite{almaraz-rivera_anomaly-based_2023}'s study highlights the importance of implementing monitoring and alerting systems to track containers' performance and resource utilization. He and Guo also implement a demo solution using an ML auto-encoder algorithm. Lastly, Yolchuyev \cite{yolchuyev_extreme_2023} proposes an eXtreme Gradient Boosting (XGBoost) based model for anomaly detection on the \acl{k8s} orchestration platform. \reviewerDc{Either way, to the best of our knowledge, no prior work has been conducted on exploiting the asynchronous design of the \ac{k8s} \ac{api} communication.}

\reviewerBc{
\textbf{Monitoring and Security in \acl{k8s}.} While the academic community has reserved little attention in the field of monitoring and security in \ac{k8s}, several tools in the market have been developed to address these issues.

OPA (Open Policy Agent)~\cite{cncf_opa_2025} is a tool that provides policy-based control for \ac{k8s}. It allows users to define policies in a high-level language and enforce them before the \ac{k8s} API server processes requests. For example, the use of the \force flag can be actively blocked using OPA. However, OPA only performs its checks before the API server, which means that container image downloads cannot be controlled in this way.

Another popular tool for monitoring \ac{k8s} environments is Falco~\cite{falco_security_falco_2025} with its plugin \texttt{k8saudit}. In addition to its runtime monitoring capabilities, Falco can ingest \ac{k8s} audit logs and generate alerts based on predefined rules. However, it works on a line-by-line basis and does not have built-in status tracking or counting features, which makes it unsuitable for tracking container image downloads. Using Falcosidekick~\cite{falco_security_falcosidekick_2025} and a visualization tool like Grafana, this limitation can be overcome, but it requires additional setup and configuration efforts that may not be feasible in all environments. Still, Falco only performs monitoring and alerting, and does not provide any mitigation capabilities.

Finally, other tools such as Tetragon~\cite{cilium_tetragon_2023} and Tracee~\cite{aqua_security_tracee_2025}, provide runtime monitoring of containerized applications. Both tools are based on eBPF and can monitor system calls, network activity, and other events in real-time. They can also detect anomalous behavior and generate alerts based on predefined rules. Like Falco, both Tetragon and Tracee perform monitoring, and Tetragon is indeed actively used in MAGI to detect system calls. However, beyond their monitoring capabilities, they no means of actively mitigating the issues they detect.
}

\section{Conclusions}
\label{sec:conclusion}

This paper demonstrates that the asynchronous implementation of some routes from the \criapi can be exploited to create a resource-based \ac{dos}. We illustrate this threat through three distinct attacks. First, we focus on resource usage, where the attack used up to 95\% of the CPU. Second, we show how this attack \reviewerBc{could interfere with the performance of other applications that run on the affected worker nodes of the cluster}. Finally, we investigate how the cache could be manipulated by the constant stream of requests performed by the attack. We further experimentally prove the effectiveness of each attack in a local testbed and on \ac{gke}. Finally, we provide a guideline to solve the problem in all the related projects and are currently collaborating with the working groups to solve this issue as soon as possible. As a stopgap, we present an eBPF-based proof-of-concept to monitor the activity in the cluster, stopping useless image downloads.

\section*{Acknowledgments}
This work was partially supported by project SERICS (PE00000014), MUR National Recovery and Resilience Plan funded by the European Union - NextGenerationEU, and by project FLUIDOS (grant agreement No 101070473), European Union’s Horizon Europe Programme.



\bibliographystyle{IEEEtran.bst}
\bibliography{kubernetes.bib}

\section{Biography Section}
 

\begin{IEEEbiography}[{\includegraphics[width=1in,height=1in,clip,keepaspectratio]{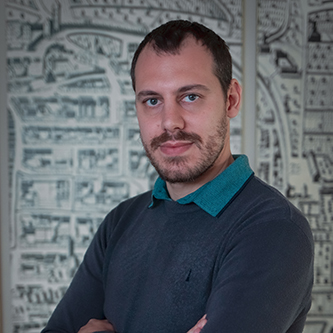}}]{Luis Augusto Dias Knob} received his MSc in Computer Science from the Federal University of Rio Grande do Sul (UFRGS), Brazil, the PhD from the Pontifical Catholic University of Rio Grande do Sul (PUCRS), Brazil, and he is a Researcher in Fondazione Bruno Kessler, Italy.
His research interests include network management, cloud computing, virtualization and containerization, and network security.
\end{IEEEbiography}
\begin{IEEEbiography}[{\includegraphics[width=1in,height=1in,clip,keepaspectratio]{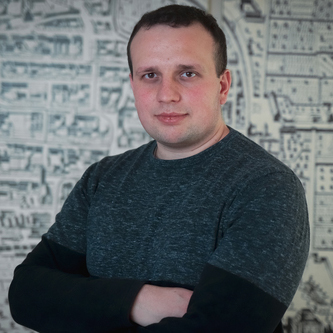}}]{Matteo Franzil} received his MSc and BSc in Computer Science from the University of Trento, Italy. Currently, he is a PhD student in University of Trento, Italy, with a grant funded by Fondazione Bruno Kessler. His research interests include network management, monitoring, and observability; virtualization and containerization; network security.
\end{IEEEbiography}

\begin{IEEEbiography}[{\includegraphics[width=1in,height=1in,clip,keepaspectratio]{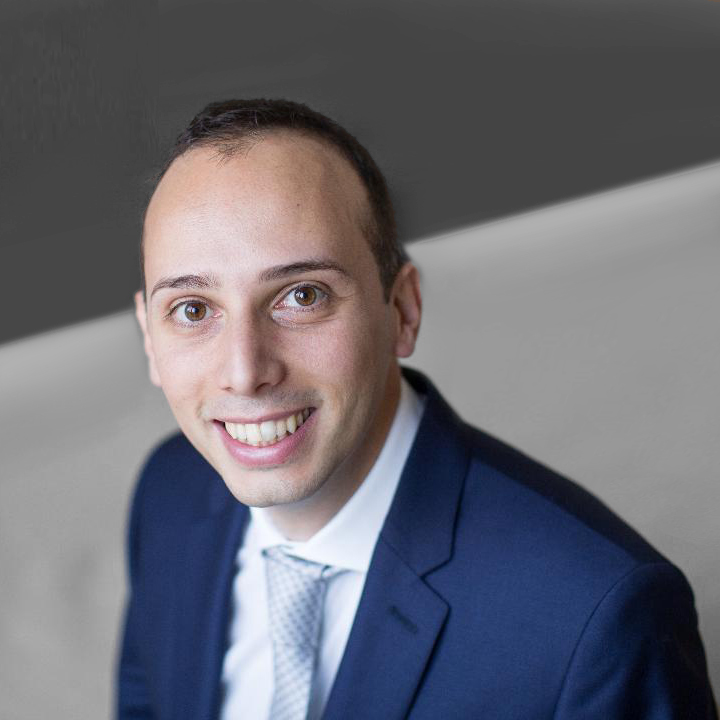}}]{Domenico Siracusa} is associate professor at the University of Trento. Previously, he was the head of the RiSING research unit at Fondazione Bruno Kessler (FBK). His research interests include infrastructure security and robustness, service orchestration and management, cloud and fog computing, and SDN/NFV and virtualization. Domenico authored more than 100 publications appeared in international peer reviewed journals and in major conferences on computing and networking technologies.
\end{IEEEbiography}


\vfill

\end{document}